\documentclass{article}

\usepackage{arxiv}

\usepackage[utf8]{inputenc} 
\usepackage[T1]{fontenc}    
\usepackage{hyperref}       
\usepackage{url}            
\usepackage{booktabs}       
\usepackage{amsfonts}       
\usepackage{nicefrac}       
\usepackage{microtype}      
\usepackage{lipsum}
 \usepackage{amsmath}
\usepackage{graphicx}
\graphicspath{ {./images/} }

\title{Clinfo.ai: An Open-Source Retrieval-Augmented Large Language Model System for Answering Medical Questions using Scientific Literature}

\author{
Alejandro Lozano \\
  Department of Biomedical Data Science\\
  Stanford University\\
   Stanford, CA, USA  \\
  \texttt{lozanoe@stanford.edu} \\
   \And
Scott L Fleming\\
 Department of Biomedical Data Science\\
  Stanford University\\
   Stanford, CA, USA  \\
  \texttt{scottyf@stanford.edu} \\
  \And
 Chia-Chun Chiang \\
 Department of Neurology \\
Mayo Clinic \\
Rochester, MN \\
  Human Centered Artificial-Intelligence Institute\\
  Stanford University\\
  Stanford, CA, USA\\
  chiang.chia-chun@mayo.edu \\
   \And
Nigam Shah \\
Center for Biomedical Informatics Research \\
Clinical Excellence Research Center \\
Stanford University \\ 
Technology and Digital Solutions \\
Stanford Healthcare \\
nigam@stanford.edu \\
}

\begin{document}
\maketitle

\begin{abstract}
The quickly-expanding nature of published medical literature makes it challenging for clinicians and researchers to keep up with and summarize recent, relevant findings in a timely manner. While several closed-source summarization tools based on large language models (LLMs)
now exist, rigorous and systematic evaluations of their outputs are lacking. Furthermore, there is a paucity of high-quality datasets and appropriate benchmark tasks with which to evaluate these tools. We address these issues with four contributions: we release Clinfo.ai, an open-source WebApp that answers clinical questions based on dynamically retrieved scientific literature; we specify an information retrieval and abstractive summarization task to evaluate the performance of such retrieval-augmented LLM systems; we release a dataset of 200 questions and corresponding answers derived from published systematic reviews, which we name PubMed Retrieval and Synthesis (PubMedRS-200); and report benchmark results for Clinfo.ai and other publicly available  OpenQA systems on PubMedRS-200. 
\end{abstract}

\keywords{Large Language Models, Abstractive Summarization, Artificial Intelligence, Clinical Medicine, Generative AI, Interactive Systems, ChatGPT} 

\vspace{5em}

\copyright\ 2023 The Authors. Open Access chapter published by World Scientific Publishing Company and distributed under the terms of the Creative Commons Attribution Non-Commercial (CC BY-NC) 4.0 License.

\newpage


The aggregation and distribution of medical knowledge, facilitated by platforms such as PubMed or Cochrane, enables healthcare professionals and medical researchers to stay abreast of the latest scientific discoveries and make informed decisions based on up-to-date scientific evidence \cite{bougioukas2020keep}. However, the staggering influx of more than 1 million papers  each year  into PubMed alone (equivalent to two papers per minute as of 2016) \cite{landhuis2016scientific} highlights the daunting task of keeping up with scientific findings \cite{van2021open}. 
This is especially true for practicing clinicians, who face the challenge of keeping track of the most updated research findings in all areas related to their patient care duties \cite{andrews2005information}.

Existing technologies fail to adequately satisfy the information needs of health care professionals and researchers. In daily  practice,  clinicians have on average one care-related question for every other patient seen \cite{del2014clinical} and they refer to sources like PubMed or UpToDate to obtain summarized information answering these questions \cite{daei2020clinical}. Questions that cannot be answered within 2 to 3 minutes are often abandoned, potentially negatively impacting patient care and outcomes  
 \cite{del2014clinical, ely2005answering}.  
 While systematic review  (SR) articles can provide quick answers to clinical questions, many questions are not answerable through existing reviews. On  the other hand, manually synthesizing findings from multiple primary sources without the help of a published review article can be extraordinarily time consuming. Review articles take on average 67.3 weeks to complete \cite{borah2017analysis}, and those written reviews may not even include the most updated research published in the literature.
Question-answering tools that leverage frequently updated external electronic resources would enable researchers and clinicians to obtain up-to-date information in a more efficient way that benefits scientific discovery and quality of patient care \cite{cook2017context,lobach2012enabling,bonis2008association,isaac2012use,reed2012relationship}.

In previous decades, applications
that integrated clinical systems with on-line
information  to answer
  users' information needs  (e.g., ``infobuttons'')  \cite{cimino1997supporting}
were typically  driven by  semantic networks. Other works such as CHiQA proposed a combination of 
knowledge-based, machine learning,  and deep learning approaches to develop a question-answering system using patient-oriented resources to answer consumer health questions \cite{demner2020consumer}. 

The new capabilities of agents powered by large language models (LLM) has accelerated the development of automated literature summarization tools. Most of these solutions tend to be privately developed, closed-source solutions based on retrieval-augmented \cite{lewis2020retrieval}  (RetA)  LLMs \cite{jin2023pubmed} (e.g. Scite \cite{nicholson2021scite}, Elicit \cite{elicit}, GlacierMD \cite{glaciermd}, Consensus \cite{consensus}, OpenEvidence \cite{openevidence}, Statpearls semantic search \cite{statpearls}). However, the paucity of publicly available technical reports describing these systems and the lack of appropriate guidelines, regulations, and evaluations to ensure their safe and responsible usage is an urgent concern \cite{healthcare11060887}.

This  Natural Language Generation (NLG) problem has been exacerbated by a lack of (1) representative datasets and associated tasks, and (2) automated metrics for evaluating RetA LLMs on said tasks.
 
Fortunately, developments in the LLM evaluation space have shown that a number of automated metrics correlate moderately with human preference, even in domain-specific scenarios (including medicine) \cite{ni2023NLG,zhong2022towards,fleming2023medalign}. 

Building on these advancements, we provide four contributions:

\vspace{1em }

\begin{enumerate}
    \item  Clinfo.ai \footnote{\href{https://github.com/som-shahlab/Clinfo.AI}{https://github.com/som-shahlab/Clinfo.AI}}, the first publicly available, open-source, end-to-end retrieval-augmented LLM-based system for querying and synthesizing the clinical literature.
    The system is hosted as a publicly available WebApp at \href{https://www.clinfo.ai/}{https://www.clinfo.ai/}.

    \item An open information retrieval and abstractive summarization task specification designed to evaluate an algorithm's ability to both retrieve relevant information and adequately synthesize it. In the task setup, both the information retrieval and abstractive summarization sub-tasks are  compared to  gold standard (human generated but pragmatically retrieved) references and answers. Furthermore, our task is defined to truly resemble RetA deployment conditions (enabling the evaluation of already deployed but potentially closed-source systems).

    \item PubMed Retrieval and Synthesis (PubMedRS-200), a publicly available dataset of 200 questions structured in Open QA format, paired with answers derived from systematic reviews and corresponding references. 

    \item Benchmark results for Clinfo.ai and other publicly available OpenQA systems on PubMedRS-200).
 
\end{enumerate}

\section{Related Work}

\textbf{LLMs in healthcare}
The remarkable performance of LLMs in the general domain has brought about a revolution in the field of natural language processing \cite{bakker2022fine}, showcasing exceptional capabilities in tasks like summarization, question-answering, and NLG \cite{bubeck2023sparks}. Given their wide utility, researchers are now actively exploring applications of LLMs in healthcare \cite{sallam2023chatgpt, eysenbach2023role, cascella2023, fleming2023assessing}. Several LLMs  have achieved human-level performance on  numerous medical professional licensing exams such as the United States Medical Licensing Exam (USMLE) \cite{kung2023performance}. Other works have demonstrated promise in various healthcare-inspired tasks, such as automated clinical note generation and reasoning about public health topics\cite{sallam2023chatgpt, eysenbach2023role, cascella2023, fleming2023assessing}. However, NLG tasks and publicly available benchmarks that directly address true medical needs are still underrepresented in the literature. Such tasks and benchmarks are especially important for estimating the capabilities and risks of LLMs in the clinical domain.

LLMs have several documented disadvantages and risks. First, updating LLMs with new knowledge and information is challenging and inefficient \cite{mitchell2022memory}.
Second, the training objective of LLMs to predict the most probable next token can cause these models to generate inaccurate information (hallucination), requiring costly and imperfect post-hoc model adjustments like reinforcement learning with human feedback (RLHF) \cite{zhao2023survey}. More importantly, most popular consumer-facing LLMs (e.g., OpenAI's GPT-4 \cite{bubeck2023sparks}, Meta's Llama 2 \cite{touvron2023llama}, Anthropic's  Claude 2 \cite{anthropic2023claude}) do not provide references pointing to their source of information, even when the model's output is factual. This can engender distrust with users in many scientific domains, including healthcare. Prior work has proposed ReTA LLMs \cite{lewis2020retrieval} to solve the information provenance issue and have shown promising results. These ReTA LLMs do not require post-hoc model editing in order to incorporate new knowledge.

\textbf{Retrieval Augmentation Question Answering LLMs in Medicine} Hiesinger et al. \cite{hiesinger2023almanac} introduced Almanac, a novel LLM integrated with a vector database and calculator, designed to answer 130 clinical questions generated by a panel of five board-certified clinicians and resident physicians. The results showed that Almanac surpassed a standard LLM (GPT-4) in factuality, safety, and correctness, indicating that retrieval systems lead to more accurate and reliable responses to clinical inquiries. Soong et al. \cite{soong2023improving} evaluated GPT-3.5 and GPT-4 models against a custom RetA LLM using a set of 19 questions. The evaluation, based solely on human judgments, revealed that both GPT-3.5 and GPT-4 exhibited more hallucinations in all 19 responses compared to the RetA model. 
While these works on RetA LLM systems represent significant progress, they suffer from at least two shortcomings: (1) they typically require human evaluation, making systematic benchmarking of new systems challenging and unscaleable; (2) they often focus solely on evaluating an LLM's output, disregarding the relevance of the information  retrieved to generate an answer. Deciding which ``relevant'' sources should be summarized can be just as challenging as generating the actual summary. Hence there is a need for a benchmark that enables integrated evaluation of both a system's ability to select relevant documents as well as its ability to summarize these documents.


\section{Materials and Methods}

\subsection{Dataset Generation}

\vspace{1em}


\begin{figure}[h]
    \centering
    \includegraphics[width=0.55\textwidth]%
    {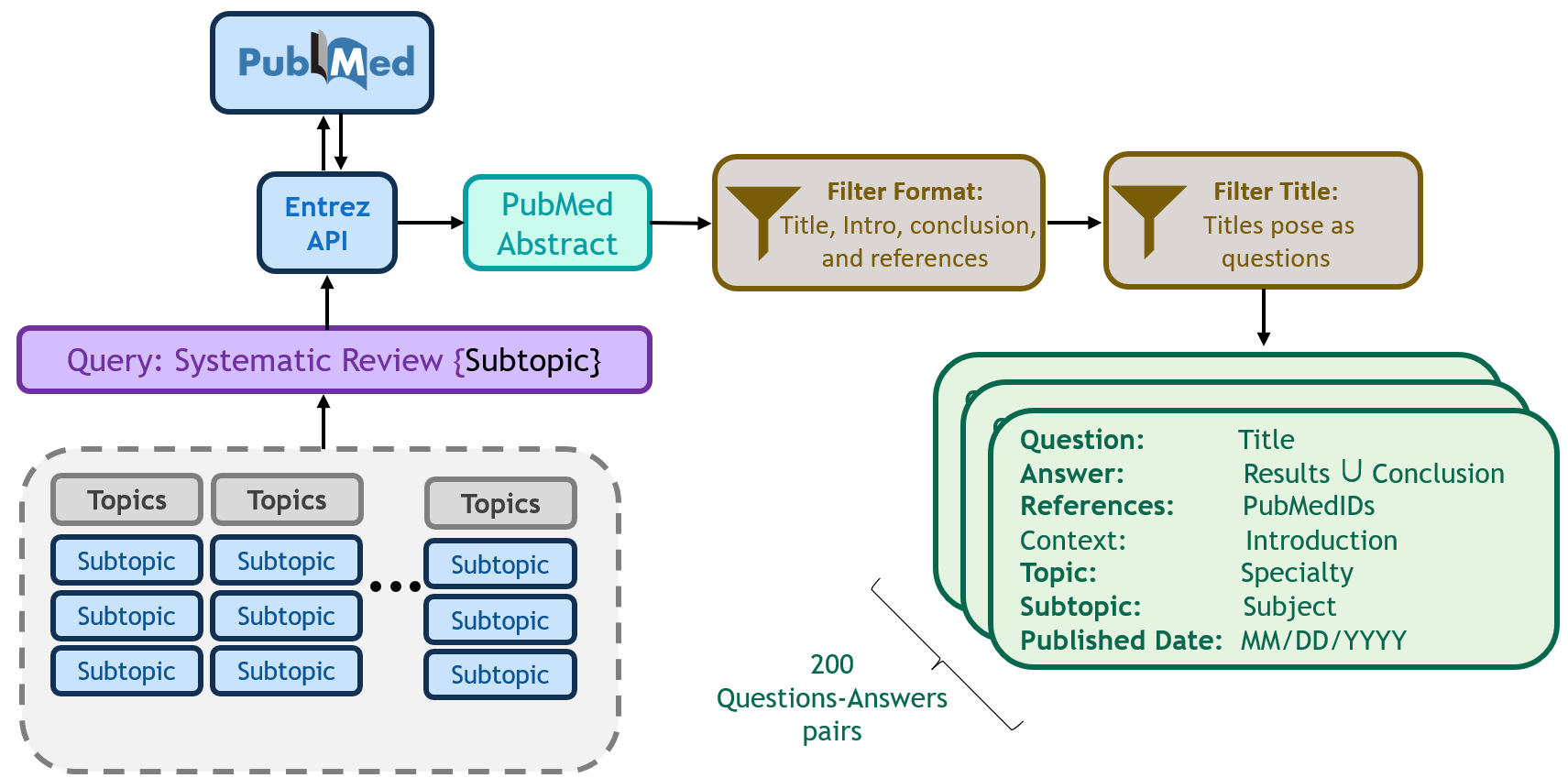}
    \caption{ Schematic Representation of the Protocol for Retrieving Abstracts from PubMed and Generating Title-Based Questions}
    \label{fig:database_protocol}
\end{figure}

 PubMed  is a  free resource supporting search and retrieval of biomedical literature \cite{sayers2023database}. As prior work has demonstrated, a large quantity of research papers available in this index  are phrased as questions, and it is possible to structure them in a question-answer format \cite{jin2019pubmedqa,scells2018generating}. Extending this idea, we created an open information retrieval and abstractive summarization dataset, using SR as a proxy for inquiries of medical interest. The rationale is that SRs are structured reviews written by human experts which summarize the pertinent literature related to a question of interest in an evidence-based manner \cite{pourreza2022towards}. In writing a SR, experienced authors (1) screen the published literature in a systematic way and include studies in a standardized manner; (2) critically evaluate methodology and reported outcomes of the included studies; and (3) carefully extract data, summarize original research findings, and in some instances, conduct additional statistical analysis of extracted results from studies including randomized controlled trials, observational cohort studies, case series and other qualitative studies on a specific topic.  Furthermore, SRs are extensively used to provide evidence for various purposes, including policy-making, clinical practice guidelines, health technology assessment, and decision making in healthcare  \cite{khan2011systematic}. As SRs unify and present a comprehensive overview of a given subject by human experts, we chose to leverage published SRs as gold standards when building our database.
 
To populate such a dataset, we employed  E-utilities, a  public API to the NCBI Entrez system \cite{sayers2023database} , to access PubMed and construct question-answer pairs with their respective references. Figure \ref{fig:database_protocol}  illustrates our process in detail. First, we established a comprehensive selection of medical specialties and subspecialties. Second, we formulated a query to retrieve Systematic Reviews relevant to each medical specialty/subspecialty. Upon constructing the specialty-specific queries and retrieving associated abstracts, we retrieved all papers structured in a format that can be easily converted to questions-answer pairs (as noted by Jin et al 2019\cite{jin2019pubmedqa})  namely Title, Introduction, Conclusion, and References. Third, we applied another filtering process, narrowing down to solely those publications whose titles included an explicit question (i.e., publications whose titles including question marks). The questions from these titles were extracted.

Finally, two human evaluators (AL and SF) manually reviewed the retrieved questions and extracted an answer to each question using minimally modified text from the results and conclusions section of the corresponding SR abstract. Concretely, in order to generate each answer, the human reviewers removed from the Results and Conclusions section of the abstract any text describing the structure or design of the systematic review (e.g., ``We used PubMed to retrieve 100 papers''), leaving only text that directly addressed the question extracted from the SR's title. In the process, abstracts that were lacking substantive results and abstracts that merely described research proposals (e.g. descriptions of future work) were entirely removed.  

 \subsection{Clinfo.ai: An LLM Chain for Information Retrieval and Synthesis} 

\begin{figure}[h]
    \centering
    \includegraphics[width=0.66\textwidth]{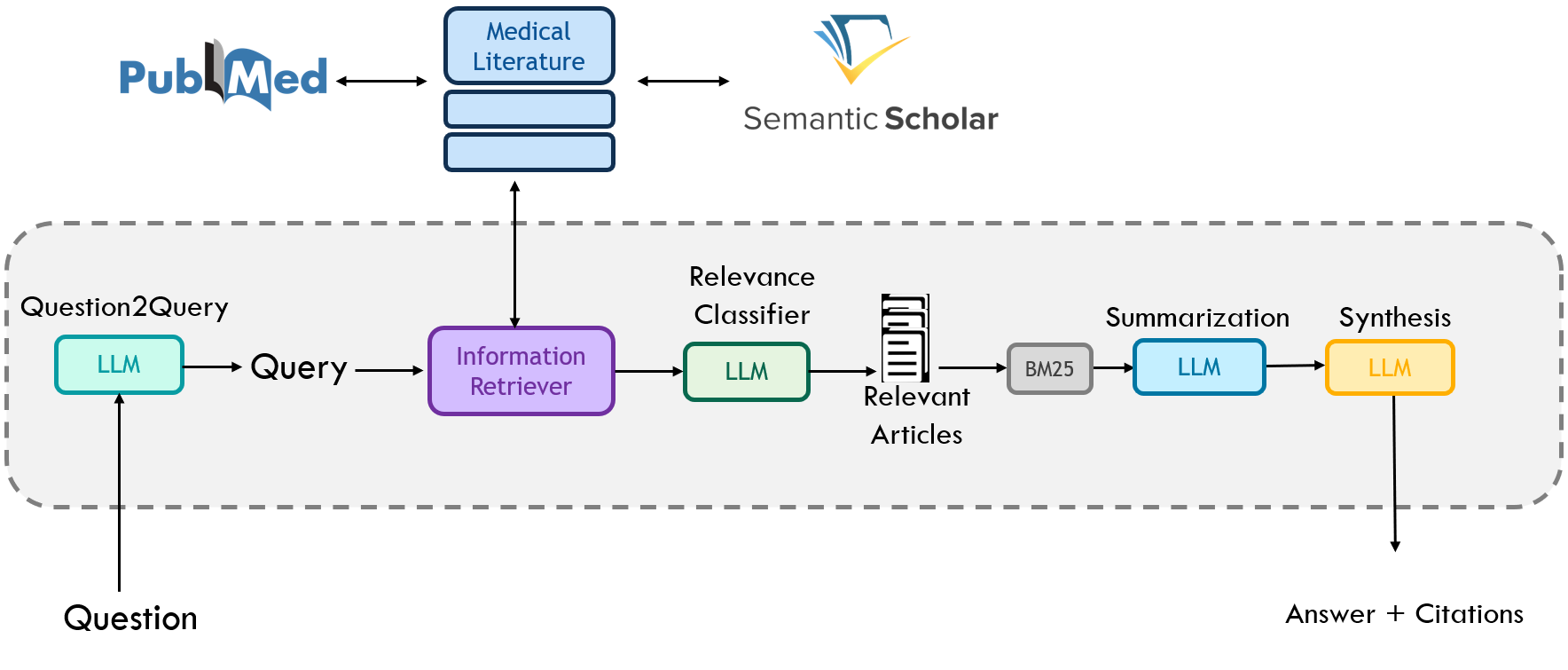}
    \caption{Clinfo.ai: A RetA LLM system for retrieving and summarizing scientific articles}
    \label{fig:clinfo}
\end{figure}

Our proposed RetA LLM system, Clinfo.ai, consists of a collection of four LLMs working conjointly (an LLM chain \cite{wu2022ai}) coupled to a Search Index (either PubMed or Semantic Scholar) as depicted in Figure \ref{fig:clinfo}.  Previous works have observed that very large language models (e.g., 100B parameters or more) exhibit zero-shot reasoning capabilities, where task-specification prompts can be used to guide the
LLM output without further fine-tuning  \cite{NEURIPS2020_1457c0d6, NEURIPS2022_8bb0d291}. We leverage the zero-shot reasoning capabilities of two LLMs, specifically OpenAI's GPT-3.5 and GPT-4 models, to complete each step in the LLM chain depicted in Figure \ref{fig:clinfo}. All prompts used in each step of the chain are available in the supplemental material \footnote{\href{https://github.com/som-shahlab/Clinfo.AI/tree/main/SupplementalMaterial}{https://github.com/som-shahlab/Clinfo.AI/tree/main/SupplementalMaterial}}. We use LangChain's API to send prompts and receive outputs from GPT-3.5 and GPT-4. While different models could technically be used through this entry point, our experiments are limited to OpenAI's GPT-3.5 and GPT-4 models (snapshots gpt-3.5-turbo-0613 , gpt-4-0613 respectively). For both models, we employ a temperature of 0.5 and a max token generator limit of 1024.

\subsubsection{Query Generator}

In our Clinfo.ai system, the input is the question submitted by the user. Once a question is submitted, the primary task of the query generator (labeled ``Question2Query'' in Figure \ref{fig:clinfo}) is to construct a PubMed (or Semantic Scholar) query that efficiently retrieves a substantial number of relevant articles pertaining to the posed question. This is achieved by instructing the model to incorporate the most crucial and relevant keywords that accurately represent the query's context and requirements.

\begin{figure}[h]
    \centering
    \includegraphics[width=0.70\textwidth]{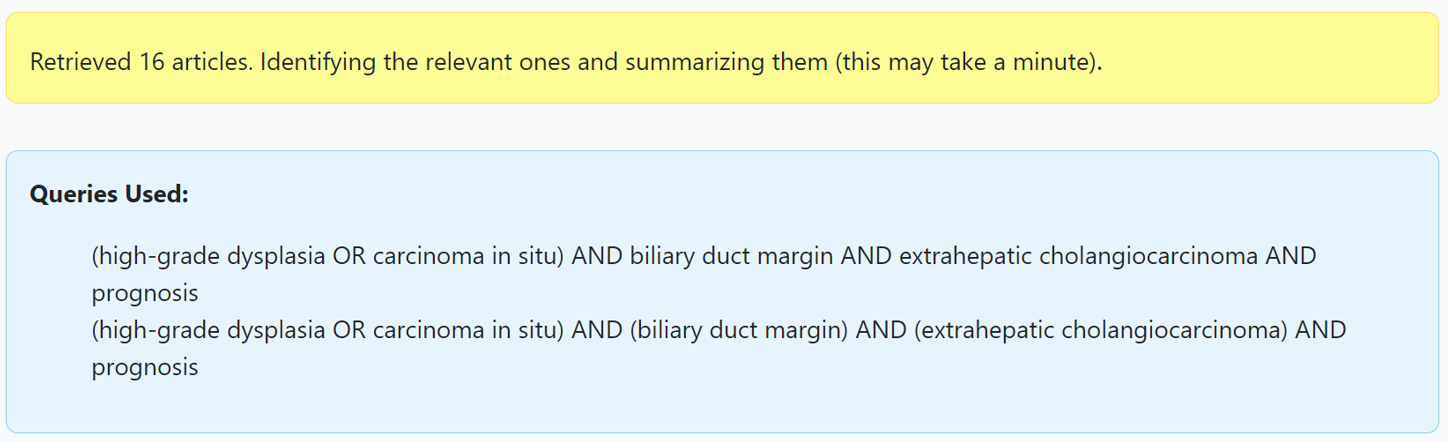}
    \caption{Query Generated by Clinfo.ai for question:  ``Does high-grade dysplasia/carcinoma in situ of the biliary duct margin affect the prognosis of extrahepatic cholangiocarcinoma?''}
    \label{fig:clinfo_query_}
\end{figure}

\vspace{-1.em}

\subsubsection{Information Retriever}
In a similar fashion to the Dataset Generation process, we utilize the Entrez API to fetch abstracts from PubMed using the output generated by the Query Generator. By leveraging the Entrez API, we are able to programmatically access and retrieve the relevant abstracts that match the constructed PubMed queries. Because LLM output is stochastic and different queries may capture different aspects of the literature, we take the union of all papers returned by three LLM-generated queries (each with the same prompt but different seeds).

\vspace{-1.em}

\subsubsection{Relevance Classifier}

Since the query generator emphasizes recall over precision (i.e., it retrieves as many potentially relevant articles as possible), it is crucial to classify the relevancy of the retrieved articles. To achieve this, we adopt an LLM-enabled binary classification approach, wherein each article is categorized as either relevant or not relevant to the posed question using GPT-3.5. Once the relevant articles are identified, we make use of the full abstract metadata of each article to construct their citations in the IEEE format. 
If more than 35 relevant articles are deemed relevant, the user can decide to re-rank and filter them using BM25 \cite{trotman2014improvements}.

\vspace{-1.em}

\subsubsection{Summarization}
The penultimate step in Clinfo.ai uses an LLM to summarize each relevant abstract within the context of the user-submitted question. 

\vspace{-1.em}

\subsubsection{Synthesis}
In the final step of Clinfo.ai, the relevant article summaries are organized as an ordered list, with each number in the list corresponding to a citation. This structured list of article summaries is then fed to a LLM with the task of constructing a concise and informative summary. The LLM is also instructed to utilize only the provided article summaries and no other additional information, relying on the structured list of citations to reference and accurately attribute each finding.

\subsection{www.clinfo.ai: A Clinfo.ai User Interface via Web Application}

\begin{figure}[h]
    \centering
    \includegraphics[width=0.60\textwidth]{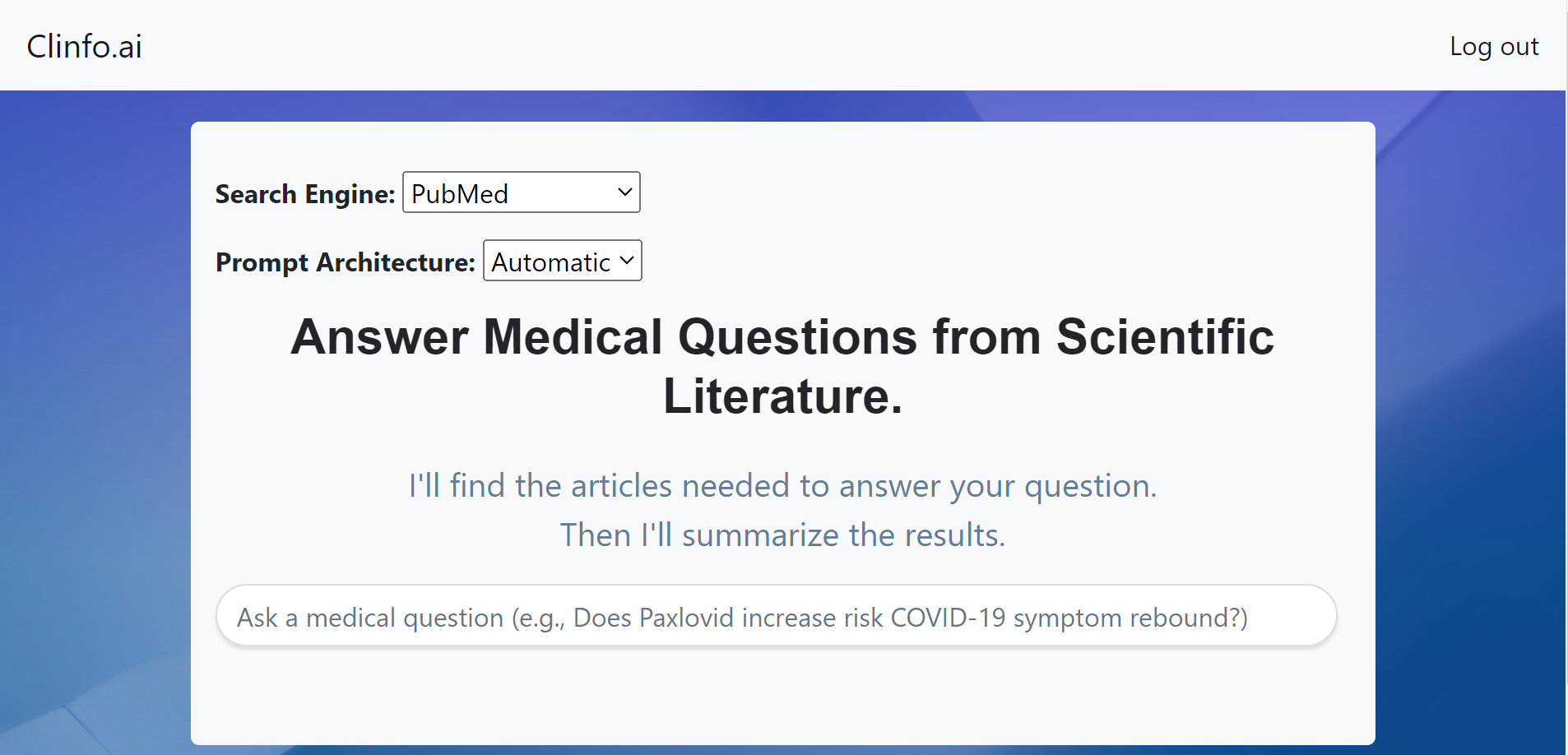}
    \caption{Clinfo.ai user interface }
    \label{fig:clinfo_interface}
\end{figure}

\begin{figure}[h]
    \centering
    \includegraphics[width=0.65\textwidth]{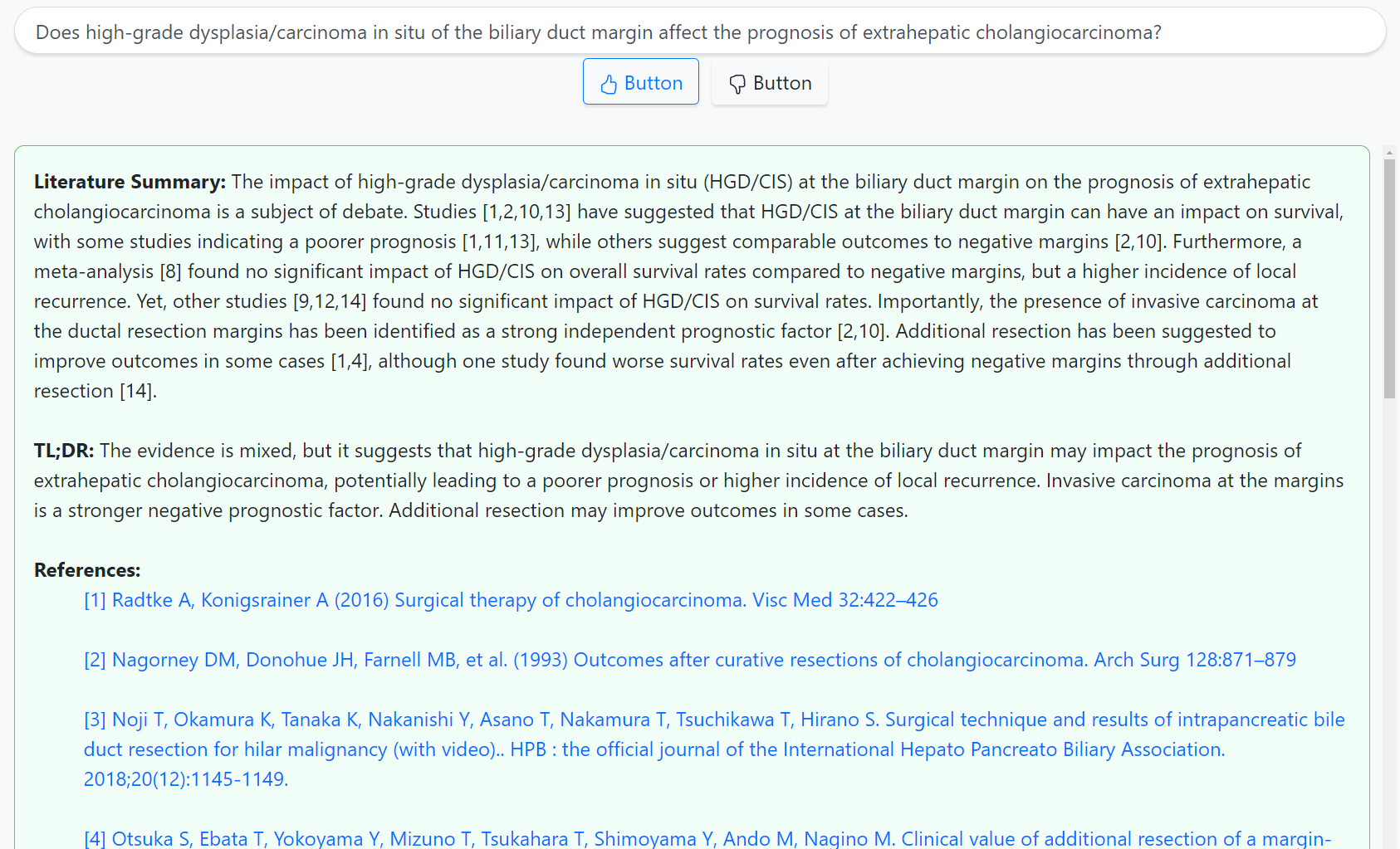}
    \caption{``Literature Summary'' (Synthesis) and ``TL;DR'' constructed with Clinfo.ai  for the question, ``Does high-grade dysplasia/carcinoma in situ of the biliary duct margin affect the prognosis of extrahepatic cholangiocarcinoma?'' (not all references are included in figure)}
    \label{fig:clinfo_re}
\end{figure}

To facilitate interaction with our system, we developed a web application that allows users to submit their own questions and/or customize the prompts. The latter enables users to tailor the system according to their individual preferences and needs, as illustrated in Figure \ref{fig:clinfo_interface}.
The entire process provides real-time access, displaying the queries generated during the search (as shown in Figure \ref{fig:clinfo_query_}), the number of retrieved articles, a concise summary of each important article, and a final ``Literature Summary'' (or ``Synthesis'', to distinguish it from the individual article summaries) accompanied by an abbreviated answer to the question (``TL;DR''). Additionally, the references are presented as hyperlinks, enabling users to verify both the validity of the reference and the information captured from it. It is possible that even after summarizing an article's abstract, Clinfo.ai may not include that article in the final Literature Summary or ``TL;DR''. Nevertheless, we ensure that all relevant articles are presented to the user so that they can access and explore them as needed. 
An example of a final Literature Review constructed with Clinfo.ai is shown in Figure \ref{fig:clinfo_re}

\subsection{Task Description and Evaluation}

The task is defined in a three step manner: 

\begin{enumerate}
    \item Given a question, generate a query to retrieve a set of articles;
    \item Given the provided articles, determine their relevancy to the question;
    \item Given relevant articles, summarize the findings.    
\end{enumerate}

Step (2) is evaluated based on precision and recall. Considering the set of all documents $D$, $RET(D,k)$ denotes  the set of $k$ retrieved documents deemed relevant and $REL(D,q)$ the set of all documents referenced by a SR. We define precision and recall in this context as follows:

\begin{eqnarray}
\text{precision} = \frac{|RET(D,k)	\cap  REL(D,q)|}{|RET(D,k)|}\label{aba:appeq2}\\ 
\text{recall} = \frac{|RET(D,k)	\cap  REL(D,q)|}{|REL(D,q)|}\label{aba:appeq3}
\end{eqnarray}

\vspace{1em}
Step (3) is conducted using both 
source-free  (SF) and source-augmented (SA) automated metrics. Source-free metrics compare a model's output to a gold standard reference summary, without including any information from the articles used to generated the gold standard summary. For our evaluation purposes, the gold standard is the human-curated answer (derived from conclusions and/or results of each SR). On the other hand, SA metrics additionally consider relevant context  to evaluate the quality of model-generated outputs.  For our experiments, context is constructed by concatenating a  SR's introduction, results, and conclusion sections. The SA metrics we employed  (and the LMs they use)
include
UniEval \cite{zhong2022towards} 
 (T5 -large), \textsc{Comet}  (XLM-RoBERTa) \cite{rei2020comet}, and  CTC  Summary Consistency (BERT) \cite{deng2021compression}.

UniEval is a multi-dimensional evaluator  designed for summarization tasks and  takes into account four key dimensions (and their corresponding overall average): 
\begin{itemize}
\item \textbf{Coherence:} Assesses whether  the summary forms a cohesive and rational body of text;
\item \textbf{Consistency:} Evaluates the factual alignment between the information presented in the summary and the content of the source document;
\item \textbf{Fluency:} Assesses the  readability and linguistic fluency of   a summary;
\item \textbf{Relevance:} Measures whether the summary contains only the important information from the source document.
\end{itemize} 

\textsc{Comet} is an evaluation metric developed to assess the
quality of Machine Translation  (MT) systems. Despite being trained on multilingual MT outputs, it performs remarkably well in monolingual settings, when predicting summarization output
quality \cite{mateusz2022comet}. CTC is an evaluation framework, based on information alignment between input, output, and  context,
for compression (e.g summary), transduction (e.g translation), and creation (e.g. conversation).

Finally we perform an evaluation using SF metrics, including
BERTScore \cite{zhang2019bertscore}, ROUGE-L \cite{lin2004rouge},  METEOR \cite{banerjee2005meteor}, chrF \cite{popovic2015chrf} , GoogleBLEU (based on \cite{papineni2002bleu}),  CTC Summary (without providing context) , and  CharacTer \cite{wang-etal-2016-character}. The majority of these metrics have  shown moderate correlation with human preference and are widely reported in NLG tasks \cite{ni2023NLG,zhong2022towards}. 

The multi-dimensional  evaluation based on source-augmented metrics makes the assumption  that an LLM+RetA model is able to (1) retrieve abstracts of works that were deemed relevant by an author of a SR and (2) synthesize them in a similar fashion. We acknowledge that if this assumption is not met, the evaluation would heavily penalize the output. Conversely, if the system retrieves an article that was not considered by a SR but bears a similar semantic meaning to an article present in the references of a SR, the evaluation would not penalize the generated text. For our proposed method, both behaviors are desired.


\section{Baselines and Experiments}
\begin{table}[h]
\centering
\caption{Performance on 146 questions from PubMedRS-200 using source-augmented (SA) metrics: UniEval (T5-large), \textsc{Comet} (XLM-RoBERTa), CTC summary (BERT)}
\resizebox{\textwidth}{!}{
\begin{tabular}{lcccccccccc}
\toprule
& \multicolumn{5}{c}{ Unified Multi-Dimensional Evaluator (UniEval)} &  & CTC (SA)  &  \\
Model & Coherence $\uparrow$ & Consistency $\uparrow$ & Fluency $\uparrow$ & Relevance $\uparrow$ & Overall $\uparrow$ & \textsc{Comet} $\uparrow$ & Consistency $\uparrow$ & Avg. Length \\
\midrule
\textbf{LLM} 
 &&&&&&&&&\\
GPT-3.5 & 0.908 (0.149)	& \underline{0.694 (0.144)}	& \underline{0.947 (0.059)}	& \underline{0.939 (0.101)}	& \underline{0.872 (0.082)} & 0.676 (0.075)	& 0.865 (0.017) & 104.834 (47.778) \\
 GPT-4 & \underline{0.915 (0.099)}	& 0.655 (0.145)	& 0.942 (0.051)	& 0.929 (0.078)	& 0.86 (0.062)	 & \underline{0.677 (0.075)}	& \underline{0.866 (0.017)} & 84.214 (39.772) \\
 \midrule
\textbf{LLM + RetA} &&&&&&&&&\\
\textit{Restricted Search} &&&&&&&&&\\

Synthesis \& TL;DR & \textbf{\underline{0.949 (0.065)}}	& 0.466 (0.105)	& 0.903 (0.104)	& \textbf{\underline{0.964 (0.053)}} & 0.82 (0.055)	& \underline{0.704 (0.055)}	& 0.84 (0.014)
 &205.579(46.181)\\
Synthesis  & 0.925 (0.066)	& 0.394 (0.11)	& 0.893 (0.119)	& 0.939 (0.101)	& 0.788 (0.059)	& 0.693 (0.057)	& 0.842 (0.015)
 & 165.814 (40.749) \\
  TL;DR  & 0.866 (0.143)	& \underline{0.787 (0.161)}	& \underline{0.954 (0.018)}	& 0.826 (0.159)	& \underline{0.858 (0.098)}	 & 0.665 (0.078)	& \underline{0.874 (0.018)}
& 38.766 (11.682) \\
&&&&&&&&&\\
\textit{Source Dropped} 
&&&&&&&&&\\
        Synthesis \& TL;DR & \underline{0.942 (0.092)}	& 0.465 (0.104)	& 0.918 (0.085)	& \underline{0.962 (0.059)}	& 0.822 (0.055) & \underline{0.706 (0.056)}	& 0.843 (0.014)
	& 204.248 (38.394) \\
  Synthesis  & 0.925 (0.066)	& 0.398 (0.112)	& 0.912 (0.096)	& 0.943 (0.055)	& 0.795 (0.055) & 0.695 (0.059)	& 0.845 (0.016)
	& 164.938 (33.221) \\
  TL;DR  & 0.829 (0.202)	& \underline{0.763 (0.197)}	& \underline{0.953 (0.029)}	& 0.796 (0.194)	& \underline{0.835(0.13)}	 & 0.672 (0.078)	& \underline{0.876 (0.017)}
 & 38.31 (10.726) \\
 &&&&&&&&&\\
\textit{Unrestricted Search} &&&&&&&&&\\
         \textit{Our Models} &&&&&&&&&\\
       Synthesis \& TL;DR & \underline{0.945 (0.064)}	& 0.539 (0.127)	& 0.912 (0.096)	& \underline{0.962 (0.059)} & 0.84 (0.052)	& \textbf{\underline{0.721 (0.055)}}	& 0.852 (0.017) & 214.338 (44.173) \\
Synthesis & 0.916 (0.092)	& 0.48 (0.142)	& 0.904 (0.098)	& 0.935 (0.069)	& 0.809 (0.06)	 & 0.712 (0.057)	& 0.855 (0.019)
& 173.379 (38.492) \\
TL;DR  & 0.896 (0.123)	& \textbf{\underline{0.81 (0.159)}}	& \textbf{\underline{0.955 (0.012)}}	& 0.857 (0.135)	& \textbf{\underline{0.88 (0.081)}}	 & 0.681 (0.072)	& \textbf{\underline{0.88 (0.016)}}
& 39.959 (11.754) \\
 &&&&&&&&&\\
  \textit{Deployed Models} &&&&&&&&&\\
    Elicit  \cite{elicit}  & 0.854 (0.136)	& 0.352 (0.147)	& 0.743 (0.151)	& 0.902 (0.117)	& 0.713 (0.085)	& 0.7 (0.066)	& 0.866 (0.017)
 & 130.566 (22.946) \\
  Statpearls SS \cite{statpearls} & 0.753 (0.225)	& 0.383 (0.129)	& 0.93 (0.053)	& 0.845 (0.159)	& 0.728 (0.112)	& 0.633 (0.075)	& 0.841 (0.016)
&  118.172 (26.603) \\
  
\bottomrule
\end{tabular}}
\label{table:unieval_all}
\end{table}

\begin{table}[h]
\centering
\caption{Performance  on 146 questions from PubMedRS-200 using source-free (SF) metrics}
\resizebox{\textwidth}{!}{
\begin{tabular}{lcccccccccc}
\toprule
Model &  BERTScore	$\uparrow$ &  ROUGE-L $\uparrow$ &  	METEOR $\uparrow$ & 	chrF	$\uparrow$ & GoogleBLEU $\uparrow$  & CTC (SF)	 $\uparrow$  & 	CharacTer $\downarrow$ & Avg. Length \\
\midrule
\textbf{LLM} &&&&&&&&&&\\
  GPT-3.5 & \underline{0.781 (0.037)}& \underline{0.165 (0.053)}	& 0.181 (0.073)	& 30.2 (10.5)	& \underline{0.077 (0.036)}	& \underline{0.575 (0.065)}	& 0.912 (0.102)
& 104.834 (47.778) \\
 GPT-4 & 0.78 (0.037)	& 0.157 (0.049)	& \underline{0.192 (0.07)}	& \underline{31.6 (9.06)}	& 0.074 (0.031)	& 0.571 (0.064)	& \underline{0.89 (0.099)} & 84.214 (39.772) \\
 \midrule
\textbf{LLM + RetA} &&&&&&&&&\\
\textit{Restricted Search} &&&&&&&&&\\

Synthesis \& TL;DR & 0.77 (0.028)	& 0.135 (0.043)	& 0.121 (0.055)	& 21.5 (9.98)	& 0.058 (0.03)	& 0.527 (0.059)	& 0.993 (0.029)	
&205.579(46.181)\\
  Synthesis & 0.773 (0.028)	& 0.141 (0.044)	& 0.133 (0.059)	& 24.3 (10.4)	& 0.063 (0.032)	& 0.533 (0.06)	& 0.976 (0.056)	
 & 165.814 (40.749)\\
 TL;DR & \underline{0.784 (0.041)}	& \underline{0.145 (0.068)}	& \underline{0.221 (0.089)}	& \underline{32.7 (7.67)}	& \underline{0.061 (0.043)}	& \underline{0.594 (0.068)}	& \underline{0.833 (0.086)}	
 & 38.766 (11.682)\\
&&&&&&&&&\\
\textit{Source Dropped} 
&&&&&&&&&\\
       Synthesis \& TL;DR & 0.773 (0.028)	& 0.136 (0.037)	& 0.119 (0.054)	& 21.4 (9.69)	& 0.057 (0.028)	& 0.53 (0.06)	& 0.989 (0.036)	& 204.248 (38.394) \\
    Synthesis  & 0.775 (0.026)	& 0.143 (0.038)	& 0.132 (0.057)	& 24.1 (9.91)	& \underline{0.061 (0.043)}	& 0.536 (0.06)	& 0.976 (0.056)	
 & 164.938 (33.221) \\
 TL;DR  & \underline{0.787 (0.041)}	& \underline{0.148 (0.064)} & \underline{0.218 (0.078)}	& \underline{33 (6.98)}	& 0.06 (0.039)	& \underline{0.6 (0.066)}	& \underline{0.83 (0.092)}	
& 38.31 (10.726) \\
 &&&&&&&&&\\
\textit{Unrestricted Search} &&&&&&&&&\\
         \textit{Our Models} &&&&&&&&&\\
       Synthesis \& TL;DR& 0.786 (0.029)	& 0.167 (0.06)	& 0.145 (0.073)	& 23.5 (11.2)	& 0.079 (0.046)	& 0.546 (0.067)	& 0.989 (0.036) & 214.338 (44.173) \\
 Synthesis & 0.789 (0.03)	& 0.178 (0.067)	& 0.164 (0.084)	& 26.7 (12)	& 0.088 (0.051) & 0.555 (0.07)	& 0.975 (0.065)	& 173.379 (38.492) \\
 TL;DR & 0.793 (0.038)	& 0.169 (0.076)	&  \textbf{\underline{0.252 (0.092)}}	&  \textbf{\underline{35.5 (7.95)}}	& 0.076 (0.049)	&  \textbf{\underline{0.61 (0.067)}}	&  \underline{\textbf{0.825 (0.094)}} & 39.959 (11.754) \\
 &&&&&&&&&\\
  \textit{Deployed Models} &&&&&&&&&\\
     Elicit  \cite{elicit}  &  \textbf{\underline{ 0.807 (0.04)}}	&  \textbf{\underline{0.218 (0.095)}}	& 0.206 (0.093)	& 31.6 (12.5)	& \underline{\textbf{0.127 (0.085)}}	& 0.596 (0.07)	& 0.938 (0.096)	 & 130.566 (22.946)  \\
    Statpearls SS \cite{statpearls}& 0.77 (0.028)	& 0.136 (0.037)	& 0.149 (0.057)	& 26.5 (9.8)	& 0.062 (0.026)	& 0.536 (0.06)	& 0.939 (0.09)	 &  118.172 (26.603) \\
  
\bottomrule
\end{tabular}}
\label{table:sf}
\end{table}

\begin{figure}[ht]
    \centering
    \includegraphics[width=1.\textwidth]{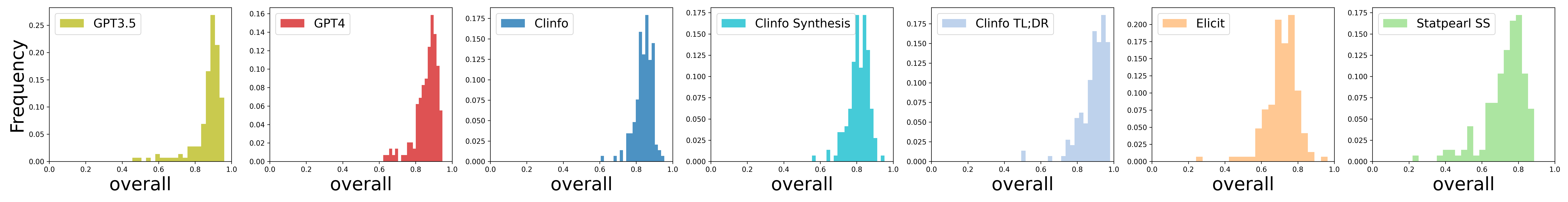}
    \caption{UniEval Overall Score of 146 questions (unconstrained by published date) from PubMedRS-200 distribution across Unrestricted Search (GPT3.5 and GPT4 zero-shot performance is added)}
    \label{fig:clinfo_query}
\end{figure}

\begin{table}[h]
\small
\centering
\caption{Clinfo.ai Precision and Recall on PubMedRS-200}
\begin{tabular}{@{}lccc@{}}
\toprule
Evaluation Regime & Precision $ \uparrow$ & Recall $ \uparrow$ & Source Included  \\
\toprule
Restricted Search & 0.224 (0.239) & 0.057 (0.061) & 0.0 (0.0) \\
Source Dropped & 0.186 (0.22) & 0.064 (0.064) & 0.0 (0.0) \\
Unrestricted Search & 0.162 (0.175) & 0.052 (0.064) & 0.965 (0.185) \\
\bottomrule

\end{tabular}
\label{table:precision_recall}
\end{table}

Using our proposed task, we evaluated the performance of GPT-4 and GPT-3.5 without retrieval augmentation, Clinfo.ai (our GPT-enabled RetA LLM system), and two deployed tools: Elicit (an AI research assistant based on LLMs, designed for facilitating literature review generation, accessed on 07-02-2023), and Statpearls Semantic Search (a free search tool for medical knowledge, accessed  on 07-25-2023).  While other automated literature summarization systems are available, at the time of this study the vast majority  require a subscription to answer multiple questions. Additionally, a subset of these systems refused to provide an answer to a significant number of the PubMedRS-200 questions as posed, making evaluation for these systems fraught and difficult to interpret. We exclude these systems from our analysis. 

Lastly, since our framework generates two outputs --- ``TL;DR'' and ``Literature Summary'' (also referred to as ``Synthesis'')  --- we conducted evaluations of three forms of Clinfo.ai's output:
(1) the synthesis of the articles retrieved and deemed relevant (``Synthesis'');
(2) the abbreviated summary distilling the proposed ``Synthesis'' into one or two sentences (``TL;DR'');
(3) the combined ``Synthesis'' and ``TL;DR''.


We recognize  that the usage of scientific literature  to extract question-answer pairs comes with the possibility that an answer deemed correct at the time of acquisition  may be incorrect as new discoveries are published.
To ensure that a system is not rewarded for simply copy-pasting the text of a retrieved source SR nor penalized when new relevant articles are published, we consider three evaluation regimes:

\begin{enumerate}
    \item  \textbf{Restricted Search (RS)}: The retrieval process is constrained to include publications up to one day before the publication date. While this approach may not guarantee the retrieval of all publications considered important by the authors of each source  systematic review, it effectively narrows down the search space to the subset of publications that could have been retrieved and deemed relevant during the review's preparation.
    
    \item \textbf{Source Dropped (SD)}: The retrieval process can retrieve articles published both before and after the source systematic review. However, if the source SR is retrieved, it is removed from the set of relevant articles and not used in the subsequent steps of the summarization process.
    
    \item \textbf{Unrestricted Search (US)} No restriction is applied; the source SR may (but need not) be included in the set of relevant articles retrieved by the system. Because we could not control the set of articles retrieved and summarized by closed-source tools like Elicit and Statpearls SS, they effectively fall within this evaluation regime.
\end{enumerate}


Finally, to ensure that conformity with the SD regime would not prevent direct comparison with the other evaluation regimes, we removed questions from all other training regimes for which Clinfo.ai could only retrieve the source article (resulting in zero articles remaining after exclusion under the SD regime). This yielded 145 SRs (80 after October 2021 and 65 before). 


\section{Experimental Results and Analysis}

\noindent\textbf{Is RetA associated with significant improvements in automated metric evaluation?}

 As reported in previous studies \cite{hiesinger2023almanac,kung2023performance,nori2023capabilities}, both GPT-3.5 and GPT-4 without RetA demonstrated strong zero-shot performance using both source-augmented (Table \ref{table:unieval_all}) and source-free  (Table \ref{table:sf}) metrics. Notably, there was no substantial performance drop observed when these models were presented with questions based on source SRs published after September 2021 (Comparing Table \ref{table:unieval_all} and Table S1 in the Supplement). While  more studies are necessary, we postulate that this  can be attributed to the models' exposure to prior published works during training. Since SRs are built upon existing literature ranging across multiple years, it is plausible that the models have been trained on relevant information that  aids them in providing accurate responses to questions based on newer research. However, comparing all LLM against LLM + RetA models, the inclusion of RetA leads to a slight improvement in the overall performance of the models when evaluated with SF and SA automated metrics, irrespective of the publication date of the source SR.  Previous works based on human evaluation have observed a similar trend, corroborating our automated evaluation framework.

\noindent\textbf{How does Clinfo.ai perform compared to other systems?}

As depicted in Table \ref{table:unieval_all}, Clinfo.ai exhibited better performance in overall UniEval compared to other RetA systems, irrespective of the chosen output strategy (Synthesis, TL;DR, or a concatenation of the two). This improvement in performance remained consistent regardless of the average length of the output, with Clinfo.ai achieving better results for both approximately 3x shorter (TL;DR) and around 2x longer outputs (Synthesis). Furthermore, this performance persisted across all different evaluation regimes, even when the source SR was dropped. This improvement amounted to at least 6.2\% and at most 14.9\% in UniEval Overall performance. These results suggest two significant points: (1) Our system is not merely copying and pasting information from an SR review. Instead, it demonstrates a genuine ability to process and present the information effectively, resulting in enhanced performance compared to other available tools; and
(2) even in the absence of a source SR, Clinfo.ai can still provide conclusions that are better aligned with a source SR's conclusion (compared to tools that might include the  source SR).


\noindent\textbf{TL;DR or Synthesis?}

Clinfo.ai TL;DR demonstrates significantly better performance compared to Synthesis and Synthesis \& TL;DR, even though they all utilize the same relevant retrieved   articles. It is worth noting that while Synthesis provides evidence to answer the question based on the retrieved articles, this evidence may not align with the original evidence reported by a Systematic Review (SR). However, the increased performance of TL;DR could be attributed to the LLM's capability to correctly identify the most salient points of the relevant articles and effectively summarize them. On the other hand, using only source-free (SF) metrics (Table \ref{table:sf}), Elicit performs better under BERTScore, ROUGE-L and GoogleBLEU, while Clinfo.ai TL;DR performs better under METEOR, chrF, CTC (SF), and CharacTer. 

These results highlight a potential limitation of automated evaluation . For instance, SF  metrics tend to reward short responses, which may not necessarily be accurate or comprehensive. On the other hand, several SA metrics can assign the best score to considerably larger generations (UniEval's Coherence and Relevance, and \textsc{Comet}), acknowledging their quality and relevance. This discrepancy in evaluation metrics raises concerns about the fair assessment of model performance and emphasizes the need for a comprehensive evaluation approach. 

Comparing different evaluation regimes, the best performance was observed under the Unrestricted Search evaluation regime, possibly due to the fact that the source SR was retrieved  on 96.5\% of the questions. As expected given the restricted set of retrievable documents, Clinfo.ai's precision was highest under the Restricted Search regime (Table \ref{table:precision_recall}).

\section{Conclusion}
The rapidly expanding medical literature and the capabilities of LLMs to process and summarize vast amounts of information have led to the development of several tools that utilize LLMs to generate on-demand summaries of published scientific literature. However, the lack of high-quality datasets and appropriate benchmarking tasks has hindered rigorous evaluations of these tools. To address this gap, we have introduced Clinfo.ai, an open-source end-to-end LLM-chain workflow designed to query, evaluate, and synthesize medical literature into concise summaries for answering questions on demand. Additionally, we introduce a unique dataset, PubMedRS-200, which consists of questions and answers extracted from systematic reviews, enabling automatic evaluation of LLM performance in Retrieval Augmentation Question Answering. Our tools and benchmarking dataset are publicly available to ensure reproducibility and to facilitate further research in harnessing LLMs for Retrieval Augmentation Question Answering tasks.

\section{Limitations}

In this study, we employed automated metrics that have demonstrated moderate-to-high correlation with human preferences, but we did not explicitly solicit human preferences to evaluate the RetA LLM systems considered. Future work should consider including human evaluation to ensure alignment of automated metrics and human preferences. 
Lastly, it is worth noting that prior studies have reported that LLMs demonstrate the ability to generate accurate Boolean operators and syntax, effectively adhering to PubMed query formats. However, our observations revealed that these models also generated hallucinated MeSH terms, which could potentially lead to the exclusion of relevant studies. To overcome this limitation, future research efforts should prioritize improving the query generation process, ensuring that generated MeSH terms are reliable and relevant for better precision and recall in medical literature search tasks.

\section{Acknowledgments}

AL is funded by Arc Institute. SF is supported by a Stanford Graduate Fellowship. This effort was supported in part by the Mark and Debra Leslie endowment for AI in Healthcare. We thank Will Haberkorn for his aid with Figure S1.

\bibliographystyle{unsrt}  
\bibliography{template}

\end{document}